\documentclass[a4paper,10pt]{article}
\usepackage{amsmath}
\usepackage{amsfonts}
\usepackage{amssymb}

\newcommand{\be}{\begin{equation}}
\newcommand{\ee}{\end{equation}}
\newcommand{\bea}{\begin{eqnarray}}
\newcommand{\eea}{\end{eqnarray}}
\newcommand{\nn}{\nonumber}

\newcommand{\la}{\lambda}
\def\Str#1{Str\left[ #1 \right]}
\def\Tr#1{Tr\left[ #1 \right]}
\def\Det#1{\mathcal{D}et\left[ #1\right]}
\def\Sdet#1{\mathcal{S}det\left[ #1\right]}
\def\SDet#1#2{\mathcal{S}det_{#2}\left[ #1\right]}
\newcommand{\bpm}{\begin{pmatrix}}
\newcommand{\epm}{\end{pmatrix}}
\def\Exp#1{{\rm Exp}\left[ #1 \right]}
\def\ID{{\bf 1}}
\newcommand{\bmm}{\begin{matrix}}
\newcommand{\emm}{\end{matrix}}
\newtheorem{theorem}{Theorem}
\newenvironment{proof}[1][Proof:]{\begin{trivlist}
\item[\hskip \labelsep {\bfseries #1}]}{\end{trivlist}}

\newcommand{\f}{\mbox{\sf f}}

\newcommand{\qed}{\nobreak \ifvmode \relax \else
      \ifdim\lastskip<1.5em \hskip-\lastskip
      \hskip1.5em plus0em minus0.5em \fi \nobreak
      \vrule height0.50em width0.5em depth0.05em\fi}

\begin{document}

\title{Equivalence of QCD in the $\epsilon$-regime and 
chiral Random Matrix Theory with or without chemical potential}

\author{~\\{\sc F.~Basile}$^{1,2}$ and {\sc G.~Akemann}$^{1,3}$
\\~\\
$^1$Department of Mathematical Sciences \& BURSt Research Centre\\
Brunel University West London\\ 
Uxbridge UB8 3PH, United Kingdom
\\~\\
$^2$Dipartimento di Fisica dell'Universit\`a di Pisa \& INFN\\
Largo B. Contecorvo 3 Ed. C, 56127 Pisa, Italy
\\~\\
$^3$Centre de Physique Th\'eorique, CNRS Luminy,\\
Case 907,
F-13288 Marseille Cedex 9, France\footnote{CPT is 
``UMR 6207 du CNRS et des universit\'es d'Aix-Marseille I, d'Aix-Marseille II
  et du Sud Toulon-Var, affili\'ee \`a la FRUMAM''}}

\maketitle 

\begin{abstract}
We prove that QCD in the $\epsilon$-regime of chiral Perturbation Theory 
is equivalent to chiral Random Matrix Theory for zero 
and both non-zero real and imaginary  chemical potential $\mu$.
To this aim we prove a theorem that relates integrals over fermionic 
and bosonic variables
to  super-Hermitian or super-Unitary groups also called superbosonisation. 
Our findings extend previous results for the equivalence of the 
partition functions, spectral
densities and the quenched two-point densities. We can show that 
all $k$-point density correlation functions agree in both theories 
for an arbitrary number of quark flavours, for either $\mu=0$ or 
$\mu\neq0$ taking real or imaginary values. 
This implies the equivalence for all individual
$k$-th eigenvalue distributions which are particularly useful 
to determine low energy constants from Lattice QCD with chiral fermions.
\end{abstract}
\vfill 
 
\begin{flushright} 
CPT-P33-2007
\end{flushright} 
\thispagestyle{empty} 

\newpage

\section{Introduction}

The application of chiral Random Matrix Theory (chRMT) to QCD, initially
introduced  
at zero chemical potential
in \cite{Shuryak:1992pi}, has led to many exact analytical results for the
Dirac operator spectrum. These have in turn found applications in 
comparison to
Lattice Gauge theory by various groups (for reviews see
\cite{VW,Akemann:2007rf}).  
The deeper reason for this analytical understanding 
of the Dirac spectrum in the phase with broken chiral symmetry can be 
traced back to the relation between chRMT and the epsilon regime of chiral
Perturbation Theory ($\epsilon\chi$PT). Being based on global symmetries
alone there are three classes of chiral symmetry breaking patterns, 
corresponding to the three
possible anti-unitary symmetries that lead to the unitary, orthogonal and
symplectic chRMT \cite{Jac3fold}. 
Here, we will only report on the first class containing QCD, which is
technically the simplest. Relatively little is known about the equivalence 
to $\epsilon\chi$PT for gauge theories in the 
two other classes, see however \cite{SmilgaVerb,toublanVerb}. 

In this paper we will show that any correlation function of Dirac operator
eigenvalues computed starting from $\epsilon\chi$PT is the same as the 
corresponding 
quantity computed from chRMT in the large-$N$ limit of infinite matrices,
corresponding to the infinite volume limit of 
$\chi$PT.
This equivalence holds both with and without chemical potential, being 
of real or imaginary type. The chRMT we use is the one introduced 
by Shuryak and Verbaarschot \cite{Shuryak:1992pi} for $\mu=0$,  by Osborn
\cite{Osborn:2004rf} for 
real  $\mu\neq0$, and by Akemann et al. \cite{ADOS} for imaginary  $\mu$
where the two latter are two-matrix models. 
The one-matrix model for real  $\mu\neq0$ introduced earlier 
by Stephanov \cite{Stephanov:1996ki} can also be mapped to
$\epsilon\chi$PT 
and is thus equivalent to \cite{Osborn:2004rf}. 
Being  
technically more involved we don't give details of our analogous 
equivalence proof here. 

The $\epsilon\chi$PT -- chRMT equivalence
is established by relating the generating functionals of
resolvents of an arbitrary finite degree, with any number of flavours $N_f$. 
In $\epsilon\chi$PT these  generating functionals are
obtained by adding additional fermionic and bosonic quarks, leading to 
supergroup integrals. 
For real  $\mu\neq0$ it is nontrivial to make them convergent, and our result 
for the corresponding  $\epsilon\chi$PT is new.
These group integrals are then shown to be equal to the 
corresponding  quantities in chRMT at large-$N$,
given by the 
expectation value of ratios of Dirac operator determinants (also called
characteristic polynomials in the chRMT language).

The technical tool we use is a theorem we prove also called superbosonisation,
which was obtained independently by 
\cite{LSZ,BEKYZ}. 
After writing the ratios of determinants in terms of superfermionic variables
and integrating out the Gaussian random matrices the theorem allows us to map 
these expressions to the supergroup integrals of $\epsilon\chi$PT.
We only take the large-$N$ limit at the very end, without performing any
saddle point approximations.
Let us emphasize that although there are many applications of RMT in Physics
\cite{GMW}, only in few
cases an exact map to the underlying microscopic theory has been achieved. 

In chRMT there are alternative ways to compute arbitrary $k$-point density
correlations,
or individual eigenvalue correlation functions, some 
without introducing resolvents. One example is the method of
orthogonal polynomials \cite{Mehta}.
On the other hand the supergroup integrals in $\epsilon\chi$PT are 
increasingly hard to calculate for $k>1$.
By establishing an equivalence for all generating functionals we can therefore 
match with all known chRMT correlation functions, were they obtained using
resolvents or not.
For $\mu=0$ all generating functionals in chRMT were computed in \cite{KJ,YG}.
For real $\mu\neq0$ all building blocks for characteristic polynomials in 
chRMT in the
large-$N$ limit follow from \cite{AOSV} and \cite{KimJac06}. 
At imaginary $\mu\neq0$ 
so far only the bi-orthogonal polynomial method has
been used in chRMT \cite{ADOS}. 

The explicit results for the $k$-point density correlation functions 
can be most conveniently read off directly from 
\cite{ADMN} for massless, and from 
\cite{DNmass,WGW} for massive correlations at $\mu=0$, for 
both massive and massless correlations from \cite{AOSV} at real $\mu\neq0$, 
and likewise from \cite{ADOS} for imaginary $\mu$. 
The corresponding individual eigenvalue
correlation functions that require the knowledge of all density correlations 
were obtained in \cite{DNW} for $\mu=0$, in \cite{ABSW} 
for real $\mu$ and 
in \cite{AD07lat} for imaginary $\mu$.

The following results were previously known about the 
$\epsilon\chi$PT -- chRMT equivalence:
partition functions with only fermions were shown to agree for zero 
\cite{Shuryak:1992pi} and non-zero $\mu$
\cite{Splittorff:2003cu,AFV,Osborn:2006xf,Akemann:2007rf}.
The equivalence for partition functions with
only bosons at $\mu=0$ follows from \cite{Fyodorov:2002ub}.
The generating functional of a $(N_f+1|1)$ supergroup integral 
leading to the spectral density was computed in 
\cite{Osborn:1998qb,Damgaard:1998xy} including $N_f$ massless fermions.
This work was 
extended to include the quenched two-point density in \cite{TV}, in 
this case too only for $\mu=0$. 
Furthermore, the quenched density at real $\mu\neq0$ was computed 
from both $\epsilon\chi$PT \cite{KJ} and chRMT \cite{Osborn:2004rf}
and found to be in agreement. 
For imaginary isospin $\mu\neq0$ the equivalence was
established up to the two-point function in \cite{Damgaard0604054,ADOS}. 
It was pointed out \cite{ADp} how in principle 
to compute the distributions of the $k$-th
individual eigenvalue from  $\epsilon\chi$PT,  using all $k$-point density
correlation functions, in order to reproduce previous 
chRMT results \cite{DNW}.
The same strategy can be applied for non-zero $\mu$ \cite{AD07lat,ABSW}. 

Our equivalence proof for all $k$-point density
correlation functions thus, in particular, 
fills this gap. Individual eigenvalue 
distributions have been used in several groups to compare with Lattice results
with exact or approximate chiral symmetry, leading to a well defined 
gauge field topology. 

A further remark concerns universality. On the chRMT side it is well known
for $\mu=0$ 
that in the microscopic large-$N$ limit the results hold for a much wider
class than Gaussian weight functions 
\cite{ADMN,KF97,DNmass}. For that reason we can restrict
ourselves to establish the equivalence for Gaussian chRMT. 
For $\mu\neq0$ less is known. 
Both the model by Stephanov \cite{Stephanov:1996ki} and 
by Osborn \cite{Osborn:2004rf} are Gaussian matrix models. As
already mentioned above the fermionic partition functions 
\cite{Splittorff:2003cu,AFV,Osborn:2006xf,Akemann:2007rf} and the quenched
density \cite{KJ,Osborn:2004rf} were shown to agree among both matrix models, 
and with $\epsilon\chi$PT. 
We can now extend this equivalence to all correlation functions. 
This agreement 
suggests universality for chRMT in the broken phase
to hold at $\mu\neq0$ as well.

Finally let us mention that for non-chiral RMT the equivalence to an
effective theory for spontaneous flavour symmetry breaking 
in 3D is simpler and has
already been fully established at $\mu=0$ by Szabo \cite{Szabo:2000qq}. We
expect 
that an extension to $\mu\neq0$ can be done following the same lines as here.

Our paper is organised as follows:
In section \ref{sec:resolvent} the resolvent method is briefly
recalled both for Hermitian operators ($\mu=0$, $\mu$ pure imaginary) and
non-Hermitian ones ($\mu$ real), in order to clarify how 
the $k$-point correlation functions are obtained from a 
partially quenched 
(pq) theory. In sections \ref{sec:herm},
\ref{sec:imaginary_chem} and \ref{sec:real_chem}  
the proofs of the equivalence for the pq theories 
for $\mu=0$, for imaginary $\mu$ and real $ \mu $ are given, respectively.
Here we write the averages over ratios of
Dirac determinants in chRMT 
as supervectors integrals, perform the chRMT average and apply our theorem
relating to $\epsilon\chi$PT supergroup integrals.
In particular the form of pq $\epsilon\chi$PT with real $\mu\neq0$
is given. 
The superbosonisation theorem is then presented and proved 
in section \ref{sec:theorem}. More technical remarks on delta functions 
and on some integrals used are deferred to appendices \ref{sec:app_delta}
and \ref{sec:app_proof_int}.

\section{The resolvent method\label{sec:resolvent}}
The resolvent method is a way to compute the $k$-point eigenvalue correlation
function of an operator distributed according to a given ensemble (in our case
these ensembles 
are QCD or chRMT). Starting from the expectation value of a ratio
of characteristic polynomials, this quantity
may be considered as the partition function of a theory with an 
additional number of fermionic and bosonic quarks, called partially quenched 
QCD (or chRMT). Both supersymmetric and replica methods rely on the resolvent
method. 

In this section
we briefly summarise it's idea for the reader's convenience, referring to the
literature for more detailed explanations \cite{Verbaarschot:2005rj}.  
Let us emphasize that the resolvent method is substantially different 
whether one considers a theory with an Hermitian or non-Hermitian operator.
QCD with zero or imaginary chemical potential is a theory with an anti-Hermitian
Dirac operator, the correspondence with an Hermitian operator is trivial. QCD  with
real chemical potential is a non-Hermitian theory.
\subparagraph{}

We start by showing the resolvent method for Hermitian theories.
The $k$-point correlation function is defined as the expectation value of the
product of $k$ $\delta$-functions: 
\begin{equation}
\rho_k(\eta_1,\dots,\eta_k)\equiv \left\langle
\prod_{j=1}^k\sum_{\lambda\in\mbox{e.v.}}\delta(\eta_j-\lambda) 
\right\rangle  \ ,
\end{equation}  
where the expectation value is computed according to the ensemble of the
operator $D$ whose eigenvalues are $\lambda$. Alternative definitions of this
quantity may be given, in order to avoid the contact terms obtained through
delta functions at coinciding arguments $\eta_i=\eta_j$ \cite{GMW}.

For simplicity we consider first the easiest case, the spectral density
($k=1$). We define its resolvent as 
\begin{equation}
G_1(z)\equiv  \left\langle \sum_{\lambda\in\mbox{e.v.}} \frac{1}{z-\lambda}
\right\rangle 
=\int d\lambda\, \rho_1(\lambda)\frac{1}{z-\lambda} \label{def_resolvent}.
\end{equation} 
Here the argument $z$ is taken to lie outside the support of $D \ \subset
\mathcal R$.
Given that the density is $z$-independent this integral equation can be
inverted as follows \cite{Muskhelishvili}: 
\begin{equation}
\rho_1(\eta)=\frac{1}{2\pi i} \lim_{\varepsilon\rightarrow 0^+}
\left[
G_1(\eta-i\varepsilon)-G_1(\eta+i\varepsilon)\right] ,
\end{equation} 
taking the discontinuity of the resolvent when crossing the support.
An essential ingredient for this result is the fact that all the
eigenvalues of an (anti-)Hermitian operator lie on the real (imaginary) axis. 
This requirement is not fulfilled by theories with a non-Hermitian Dirac
operator, invalidating these equations. 
The next step is to generate the resolvent (\ref{def_resolvent}) as a ratio of
determinants: 
\begin{equation}
G_1(z)=\left\langle \Tr{\frac{1}{z-D}}\right\rangle= 
\frac{\partial}{\partial z^\prime}\left\langle
\frac{\Det{D-z^\prime}}{\Det{D-z}}\right\rangle\Big|_{z^\prime=z} \ .
\label{G1gen}
\end{equation}  

The generalisation to the $k$-point correlation function is conceptually not
different, with a bit more attention needed in 
subtracting eventual higher order
poles in $z$ in $\left\langle \prod_{j=1}^k\sum_{\lambda\,\in\,\mbox{e.v.}}
\frac{1}{z_j-\lambda}\right\rangle$.
It can be proved that the $k$-point
correlation function may be generated through the theory with $k$ additional
bosonic and fermionic quarks. The corresponding resolvents are defined as 
\begin{equation}
G_k(z_1,\dots,z_k)
\equiv \left\langle
\prod_{j=1}^k\sum_{\lambda\in\mbox{e.v.}}
 \frac{1}{z_j-\lambda}
\right\rangle  
=\int\prod_{j=1}^kd\la_j \frac{1}{z_j-\lambda_{j}}
\rho_k(\la_1,\dots,\la_k)\ .
\end{equation}  
and can be generated as in eq. (\ref{G1gen})
\begin{equation}
G_k(z_1,\dots,z_k)=\left\langle \prod_{j=1}^k\Tr{\frac{1}{z_j-D}}
\right\rangle= 
\left(\prod_{j=1}^k\frac{\partial}{\partial z_j^\prime}\right)\left\langle
\prod_{j=1}^k\frac{\Det{D-z_j^\prime}}{\Det{D-z_j}}\right\rangle
\Big|_{z^\prime_j=z_j\,\forall j} 
\label{Gkgen}
\end{equation}  
The density correlations are then given by the discontinuities with respect to
all arguments (see e.g. \cite{Szabo:2000qq}): 
\begin{equation}
\rho_k(\eta_1,\dots,\eta_k)=\frac{1}{\left( 2\pi i\right)^k}
\lim_{\varepsilon\rightarrow 0^+} 
\sum_{\{\sigma\},\sigma_j=\pm 1 } \left(\prod_{j=1}^k\sigma_j\right)\,
G_k\left(\eta_1-i\sigma_1 \varepsilon,\, \dots,\, \eta_k -i\sigma_k
\varepsilon\right)\ .
\end{equation}  
 \subparagraph{} 

In non-Hermitian theories eigenvalues are no more localised along the real
(imaginary) axis, and, as a consequence, the correlation functions cannot be
computed through the discontinuity of the resolvent along the support of the
eigenvalues. The equation below will be used to generate the two-dimensional
$\delta$-function in the complex plane instead: 
\begin{equation}
\partial_{z^*} \frac{1}{z}=\pi \delta_{\mathbb C}^2(z)\ .
\label{def-delta2}
\end{equation} 
Because the average in eq. (\ref{def_resolvent}) is now over complex
eigenvalues the resolvent is well-defined everywhere in the complex plane,
both in- and outside the support of $D$. 
From eq. (\ref{def-delta2}) it follows that for the density we have
\begin{equation}
\rho_1(z)=\frac{1}{\pi}\partial_{z^*} G_1(z)\ .
\end{equation} 
In other words the resolvent is holomorphic outside the support, and
non-holomorphic inside.
As before the resolvent may be generated by 
differentiating the expectation value of a ratio of determinants
as given in eq. (\ref{G1gen}). 
However, this case has an additional problem due to the
fact that the microscopic field theory (such as $\epsilon\chi$PT) 
requires the introduction of additional bosonic and fermionic conjugated 
quarks. This phenomenon is due to the 
Hermiticity requirement \cite{Toublan:1999hx,Splittorff:2003cu,KimJac06} 
for bosonic field theories to be written in terms of convergent integrals, and
we will discuss that in more detail in sect. \ref{sec:bosonic_z_mu}. 

As before the above can be generalised to any $k$-point function, leading
to the introduction of $k$ additional bosonic and fermionic quarks (or to $k$
couples of conjugated bosonic and fermionic quarks, if one needs to consider
the microscopic theory), with 
\begin{equation}
\rho_k(z_1,\ldots,z_k)=\left(\prod_{j=1}^k\frac{1}{\pi}\partial_{z^*_j} \right)
G_k(z_1,\ldots,z_k)\ .
\end{equation} 
\medskip

Let us mention 
an alternative method to generate all density correlations - apart from
non-resolvent related ones: it generates the 
two-dimensional $\delta$-function using the identity
\begin{equation}
\delta_{\mathbb C}^2(z)=\lim_{\kappa\to0} \frac{k^2}{\pi (z\cdot z^*+k^2)^2}
\end{equation} 
and hence when considering the sum of the delta functions 
over all the eigenvalues $\lambda$ of a 
non-Hermitian matrix $D$ we have (see \cite{FS96}):
\begin{equation}
\sum_{\lambda\in\mbox{e.v.}}\delta_{\mathbb C}^2(z-\lambda)=
\lim_{\kappa\to0}\frac{1}{\pi} 
\Tr{\frac{\kappa^2}{((z-D)(z-D)^\dagger+\kappa^2)^2}}\ .
\end{equation} 
\medskip

\section{Zero chemical potential\label{sec:herm}}
We start from the simplest case: the Hermitian chRMT corresponding to QCD with
zero chemical potential. In principle this theory could be seen as a particular
case of the imaginary $ \mu $ case studied in sect. \ref{sec:imaginary_chem},
setting $\mu=0$.
Since computations are easier here, we present this case separately for
pedagogic reasons.

The idea of our approach to prove the equivalence of all spectral
properties in the two theories is the following: in both theories the
resolvent method allows in principle to compute all $k$-point correlation
function. So without taking the discontinuities, or anti-holomorphic
derivatives, it is sufficient to prove that, up to an irrelevant constant, the
integrals describing these generating functions of the resolvents coincide. 
From this knowledge follows that all the k-point correlation
functions are equal, and, hence all spectral properties agree.
In order to actually compute any given correlation function we may thus choose
either theory, and within that theory we may even use any other equivalent
method that is simpler. 

\subparagraph{} 
Hermitian chRMT is a very efficient way to describe the spectral 
properties of QCD without chemical potential \cite{Shuryak:1992pi,VW}. 
The equivalent of the QCD Dirac operator in chRMT is given by
\begin{equation}
{\mathcal D}_f
+m_f\equiv \bpm m_f \ID_{N_+} & iA \\ iA^\dagger & m_f\ID_{N_-}\epm
\ ,\ \ f=1,\ldots,N_f\ ,\
\end{equation} 
where $A$ is a complex $N_+\times N_-$ random matrix. It describes a fixed
topological sector of QCD with a topological charge $\nu = N_+-N_-$, where 
$ \nu $
will be kept fixed while considering the $N_++N_-\equiv N\rightarrow\infty$
limit.  

Thanks to the universality of the microscopic limit of chRMT 
\cite{ADMN,KF97,DNmass} we can choose a Gaussian weight function for
the random matrix $A$. Since we are interested in finding the spectrum and the
$k$-point correlation function using the resolvent method
\cite{Verbaarschot:2004gj,Osborn:1998qb,Damgaard:1998xy,Szabo:2000qq},
according to what was said
in sect. \ref{sec:resolvent} we consider the more general theory with $n_f$
fermionic quarks and $n_b$ bosonic ones and compute it's partition function:  
\begin{equation}
\label{pq_real}
Z_{pq}
=\left\langle \dfrac{\prod_f^{n_f}\Det{{\mathcal D}_f+m_f}}
{\prod_b^{n_b}\Det{{\mathcal D}_b+m_b}}  \right\rangle
=\int dA\, e^{-\sigma N \Tr{A^\dagger A}} \frac{\prod_{f}^{n_f}\Det{\bmm m_f
    \ID_{N_+} & iA \\ iA^\dagger & m_f\ID_{N_-}\emm}} 
{\prod_{b}^{n_b}\Det{\bmm m_b \ID_{N_+} & iA \\ iA^\dagger &
    m_b\ID_{N_-}\emm}}\ .
\end{equation}
Here $ dA $ is the flat measure in the independent entries of the complex
matrix $ A $. The quantity defined above is usually called partially-quenched
(pq) partition function.  

In the following it will be required that $Re(m_b)>0 \, \forall\, b$ 
in order to have convergent integrals  generating the
resolvents in the upper half-plane. 
The result for $Re(m_b)<0$ may
be easily recovered from the equation above\footnote{In the literature 
often a diagonal matrix $S$ with elements $s_i=\pm1$ is 
introduced to be able to work 
in the whole complex plane at once. For simplicity we omitted such a
notation.}.

At this point we follow the common procedure \cite{GMW} 
of writing the determinants in
the numerator in terms of Gaussian fermionic integrals, and the ones in the
denominator in terms of bosonic ones.
We introduce two sets of $N_+$ and $N_-$ complex-supervectors\footnote{The
  anticommuting number part of supervectors and its conjugate are independent
  Grassmann variables.} in $(n_b|n_f)$.
In this paper we will use the
  boson-fermion convention for ordering elements in supervectors.
We write
these vectors in a matrix form $\psi_{g,\alpha}, \phi_{g,\beta},
\psi^*_{g,\alpha}$ and $ \phi^*_{g,\beta}$ where Latin indices run over the
$(n_b|n_f)$ superflavours and Greek indices run over the $N_+$ ($N_-$)
eigenvalues,
\be
Z_{pq}
=\int dA\, d(\psi,\psi^*,\phi,\phi^*)\, e^{-\sigma N \Tr{A^\dagger A}}
\Exp{-\sum_{g=-n_f}^{n_b}  
\bpm \psi^{*}_{g,\alpha} \\ \phi^{*}_{g,\beta}\epm 
\bpm  m_g\ID_{\alpha,\alpha^\prime} &  i A_{\alpha,\beta^\prime} \\
i A^\dagger_{\beta,\alpha^\prime} &  m_g\ID_{\beta,\beta^\prime}\epm 
\bpm \psi_{g,\alpha^\prime} \\ \phi_{g,\beta^\prime}\epm}. 
\ee
Here $d(\psi,\psi^*,\phi,\phi^*)$ is a shorthand notation for the product of
the flat measures of the independent entries of the supervectors.  
The integral above depends on the random matrices only in a Gaussian
way: 
\begin{equation}
\Exp{-\sigma N \ A_{\alpha,\beta}A_{\alpha,\beta}^* -i A_{\alpha,\beta}
  \left( \psi_{g\alpha}^*\phi_{g,\beta}\right) -iA_{\alpha,\beta}^* \left(
  \phi_{g,\beta}^* \psi_{g\alpha}\right) } \ .
\end{equation} 
We can thus perform the Gaussian integration by completing the squares:
\begin{eqnarray}
Z_{pq} &\propto& \int d(\psi,\psi^*,\phi,\phi^*) \ \Exp{-\Str{m_g
    \cdot\sum_{\alpha}\psi_{g,\alpha}\otimes\psi_{g,\alpha}^\dagger +m_g \cdot
    \sum_\beta\phi_{g,\beta}\otimes\phi_{g,\beta}^\dagger}} \nonumber \\ 
&&\times\ \Exp{-\frac{1}{\sigma N}
\Str{\sum_\beta \phi_{g,\beta}\otimes \phi_{h,\beta}^\dagger \cdot
    \sum_\alpha\psi_{h,\alpha}\otimes\psi_{g,\alpha}^\dagger}} \ .
\end{eqnarray}
In the following we introduce the matrix notation
$M_{gh}=N\cdot\delta_{gh}m_g$, anticipating the correct scaling later 
in the large-$N$ limit to obtain $N$-independent quantities 
in the thermodynamic limit \cite{Leutwyler:1992yt}. 
This expression is a Gaussian integral in any of the two sets of
supervectors. We can easily perform one supervector Gaussian integration
($\phi,\phi^\dagger$) obtaining a superdeterminant as a result:   
\begin{equation}
Z_{pq} \propto\int d(\psi,\psi^*) \,
e^{-\Str{m_g\cdot\sum_{\alpha}\psi_{g,\alpha}\otimes\psi_{g,\alpha}^\dagger}}
\ \Sdet{\frac{1}{N}M_{gh}+\frac{1}{\sigma N} \sum_{\alpha}
  \psi_{g,\alpha}\otimes \psi_{h,\alpha}^\dagger  }^{-N_-}. 
\end{equation}

The equation above depends on the supervectors only trough the sum of external
products. We can now apply the superbosonisation theorem in eq. (\ref{theorem_eq}),  
\begin{equation} 
\frac{1}{N_+} \sum_{\alpha=1}^{N_+} \psi_{g,\alpha}\otimes
\psi_{h,\alpha}^\dagger 
\rightarrow U_{gh}\, \in \, \hat{Gl}(n_b|n_f) 
\end{equation} 
where $\hat{Gl}(n_b|n_f)$ is the maximal Riemannian submanifold of the linear
group in the $(n_b|n_f)$ superspace \cite{Zirnbauer:1998}; matrix
representations of this manifold are possible, their elements are
supermatrices with $Gl(n_b)/U(n_b) \oplus U(n_f)$ as a base
manifold\footnote{The bodies, that are the complex number parts of
the commuting numbers, of the boson-boson block and fermion-fermion
  block belong to a manifold called base manifold.} and $2 \, n_b\cdot n_f$
Grassmann variables. The Haar measure on this manifold is called $d\mu_H(a)$
and may be expressed in terms of usual integrations trough (see
eq. (\ref{claim_equivalence})): 
\begin{equation}
\int_{\hat{Gl}(n_b|n_f)}d\mu_H\, =\int_{H_1=H_1^\dagger} dH_1 \theta(H_1)
\int_{U(n_f)} d\mu_U(H_2) \, \Det{H_2}^{n_f}  
\int d\Theta\, d\Theta^\dagger\,   \Sdet{\bmm H_1 & \Theta^\dagger \\ \Theta &
  H_2 \emm}^{n_f-n_b}  \ .
\label{Haar_measure_Gl}
\end{equation} 
Here, $dH_1$ is the flat measure on the Hermitian matrices, $\theta(H_1)$ is
the product of the step function in the eigenvalues ($\theta(H_1)>0\
\Longleftrightarrow$ $H_1$ is positive definite), $d\mu_U$ is the Haar measure
on unitary matrices $H_2$ and $d\Theta,d\Theta^\dagger$ is the flat 
Grassmannian integration in the independent entries of the boson-fermion
block\footnote{The Grassmann variables $\Theta_{\alpha,b}$ and
  $\Theta^\dagger_{b,\alpha}$ are independent real Grassmann
  variables.}. Other parametrisations of $\hat{Gl}(n_b|n_f)$ have been provided
for some specific values of $(n_b|n_f)$
\cite{Damgaard:1998xy,Andreev:1994,TV}.  

The result is ($N_+=\nu+N_-$):
\begin{eqnarray}
Z_{pq}&\propto &\int_{\hat{Gl}(n_b |n_f)} d\mu_H(U)\, \ \Sdet{U}^{N_+}
e^{-\Str{\frac{N_+}{N} MU}} \Sdet{\sigma  \frac{M}{N_+}+ U }^{-N_-} \nonumber
\\  
&=& \int_{\hat{Gl}(n_b |n_f)} d\mu_H(U)\,  \Sdet{U}^{\nu}
e^{-\Str{\frac{N_+}{N}MU}}  
\Sdet{1+\sigma  \frac{1}{N_+} M\cdot U^{-1} }^{-N_-}.
\end{eqnarray}
All the equations above hold for finite $N$. 
As a last step we can perform the $N\rightarrow \infty$ limit. Here we send
the masses $m_h$ to zero while keeping $M_{gh}$ fixed. Therefore the
$N\rightarrow 
\infty$ merely leads to an expansion of the superdeterminant, 
without the need
to perform any other approximation:
\begin{eqnarray}
\lim_{N\to\infty}
Z_{pq}&=& 
\int_{\hat{Gl}(n_b |n_f)} d\mu_H(U)\, \Sdet{U}^{\nu}
e^{\Str{-\frac{1}{2}MU-\sigma M\cdot U^{-1}}} \nonumber \\ 
&\propto& \int_{\hat{Gl}(n_b |n_f)} d\mu_H(U)\, \Sdet{U}^{\nu}
e^{-\sqrt{\frac{\sigma}{2}}\Str{M\cdot U+ M\cdot U^{-1}}}\ .
\label{hermitean_final_1}  
\end{eqnarray} 
This equation is equivalent to the 
partially-quenched partition function e.g. in
\cite{Damgaard:1998xy}, after matching parameters
\be 
m_i N \sqrt{\frac{\sigma}{2}}=m_i \frac{\Sigma V}{2} \ .
\label{mass}
\ee
We denote by $V$ the volume in $\chi$PT 
and by $\Sigma$ the chiral condensate, the first low energy constant in 
$\chi$PT.
Applying the resolvent method to both
eq. (\ref{pq_real}) in the limit $N\to\infty$, 
and eq. (\ref{hermitean_final_1}) while keeping eq. (\ref{mass}) finite, 
we obtain the claimed $\epsilon\chi$PT-chRMT equivalence at $\mu=0$ of
all $k$-point correlation functions in the microscopic limit.

\section{Imaginary chemical potential\label{sec:imaginary_chem}}
The simplest generalisation of the arguments above is obtained by adding
imaginary chemical potentials. This is a Hermitian theory too and, hence, the
resolvent method may be applied in the very same way as before. The only
difference with the former section lies in a different random matrix 
Dirac operator, containing one more random matrix. 
This model introduced in \cite{ADOS} follows the idea of \cite{Osborn:2004rf}
in 
assuming that the chemical potential term in non-diagonal in matrix space. 
This apparent complication by adding more random variables in fact makes the
model simpler: for two different chemical potentials one can go 
to an eigenvalue basis and use bi-orthogonal polynomials.
Below we don't need to diagonalise the matrices as we explicitly integrate them
out. The equivalence between chRMT and $\epsilon\chi$PT 
we show thus holds for any number of different chemical potentials. However,
the integrals and density correlations are so far known explicitly 
only when having two different chemical potentials for any number of flavours.

In principle one could also follows the idea of Stephanov 
\cite{Stephanov:1996ki} with only
one random matrix and the $\mu$-term proportional to unity. 
However, for technical reasons the computation is more involved and although
we don't show it here it leads to the same result.

The Dirac operator for a quark with mass $m_f$ and imaginary chemical
potential $\mu_f$ is given by \cite{ADOS}: 
\begin{equation}
{\mathcal D}_f+m_f\equiv \bpm m_f \ID_{N_+} & iA+i\mu_f B \\ iA^\dagger +i\mu_f
B^\dagger & m_f\ID_{N_-}\epm \ ,
\end{equation}
and likewise for boson masses  $m_b$ and chemical
potential $\mu_b$.
Here $A$ and $B$ are complex $N_+\times N_-$ random matrices with Gaussian
weights. The issue of universality is more subtle here because the matrices $A$
and $B$ will couple after changing variables. We refer to \cite{ADOS} for a
more detailed discussion.

We will prove the equivalence of the partition functions for a generic theory
with $ n_b $ bosons and $ n_f $ fermions. The equivalence of the spectra will
then follow, 
applying the resolvent method. As before we will consider only bosonic
masses with positive real parts,
\begin{equation}
Z_{pq}
=\left\langle \dfrac{\prod_f^{n_f}\Det{{\mathcal D}_f+m_f}}
{\prod_b^{n_b}\Det{{\mathcal D}_b+m_b}}  \right\rangle
=\int dA\, dB\, e^{-\sigma N \Tr{AA^\dagger +BB^\dagger}}
\frac{\prod_{f}^{n_f}\Det{\bmm m_f
    \ID_{N_+} & iA+i\mu_f B \\ iA^\dagger+i\mu_f B^\dag & m_f\ID_{N_-}\emm}} 
{\prod_{b}^{n_b}\Det{\bmm m_b \ID_{N_+} & iA +i\mu_b B
\\ iA^\dagger+i\mu_f B^\dag & m_b\ID_{N_-}\emm}}\ .
\end{equation} 

From this point on, most of the steps are equivalent to the ones performed in
the previous section: we introduce two sets of $N_+$ and $N_-$
complex-supervectors in $(n_b|n_f)$ to write the ratio of determinants as a
Gaussian integral
\begin{eqnarray} 
Z_{pq}&=&
\int dA\ dB\ d(\psi,\psi^*,\phi,\phi^*) \Exp{-\sigma N
  \left(\Tr{A^\dagger A}+\Tr{B^\dagger B}\right)}  \\ 
&&\times\ \Exp{-\sum_{g=-n_f}^{n_b} 
\bpm \psi^{*}_{g,\alpha} \\ \phi^{*}_{g,\beta}\epm 
\bpm  m_g\ID_{\alpha,\alpha^\prime} &  \left(i A+i \mu_g
  B\right)_{\alpha,\beta^\prime} \\ 
\left(i A^\dagger+i \mu_g B^\dagger\right)_{\beta,\alpha^\prime} &
  m_g\ID_{\beta,\beta^\prime}\epm  
\bpm \psi_{g,\alpha^\prime} \\ \phi_{g,\beta^\prime}\epm}. \nonumber
\end{eqnarray} 
We reduce all the dependence on random matrices in terms of Gaussian functions:
\begin{eqnarray}
&&\Exp{-\sigma N \ A_{\alpha,\beta}A_{\alpha,\beta}^* -i A_{\alpha,\beta}
    \left( \psi_{g\alpha}^*\phi_{g,\beta} \right)-iA_{\alpha,\beta}^* \left(
    \phi_{g,\beta}^* \psi_{g\alpha}\right) }\\ 
&&\times \ \Exp{-\sigma N \ B_{\alpha,\beta}B_{\alpha,\beta}^* -i \mu_g
    B_{\alpha,\beta}  \left( \psi_{g\alpha}^*\phi_{g,\beta} \right) -i \mu_g
    B_{\alpha,\beta}^* \left( \phi_{g,\beta}^* \psi_{g\alpha}\right) }\ .
    \nonumber 
\end{eqnarray} 
Next we can perform the Gaussian integration completing the squares:
\begin{eqnarray}
Z_{pq} &\propto& \int d(\psi,\psi^*,\phi,\phi^*) \ \Exp{-\Str{m_g
    \cdot\sum_{\alpha}\psi_{g,\alpha}\otimes\psi_{g,\alpha}^\dagger +m_g \cdot
    \sum_\beta\phi_{g,\beta}\otimes\phi_{g,\beta}^\dagger}} \nonumber \\ 
&&\times \ \Exp{-\frac{1}{\sigma N}\Str{\sum_\beta \phi_{g,\beta}\otimes
    \phi_{h,\beta}^\dagger \cdot
    \sum_\alpha\psi_{h,\alpha}\otimes\psi_{g,\alpha}^\dagger}}
\nonumber \\ 
&&\times \ 
\Exp{-\frac{1}{\sigma N}\Str{\mu_g \cdot \sum_\beta
    \phi_{g,\beta}\otimes \phi_{h,\beta}^\dagger \cdot \mu_h \cdot
    \sum_\alpha\psi_{h,\alpha}\otimes\psi_{g,\alpha}^\dagger}}. 
\end{eqnarray} 
Here we introduce again the matrix notation 
$M_{gh}\equiv  N\cdot \delta_{gh} m_g$, as well as
$B^{(\mu)}_{gh}= \sqrt{\frac{N}{2}} \delta_{gh} \mu_g$, anticipating also the
proper scaling of $\mu$ with $N$ below.
This expression is a Gaussian integral in any of the two sets of
supervectors. As before we explicitly integrate one set of supervectors and
express the remaining external product in terms of an integration over $
\hat{Gl}(n_b|n_f) $. The result is: 
\begin{equation}
Z_{pq}\propto \int d(\psi,\psi^*) 
e^{-\Str{\frac{1}{N}M\cdot\sum_\alpha
    \psi_\alpha\otimes\psi_\alpha^\dagger}} 
\Sdet{\frac{1}{N}M+\frac{1}{\sigma N}\sum_\alpha
  \psi_\alpha\otimes\psi_\alpha^\dagger + \frac{2}{\sigma N^2} B^{(\mu)} \cdot
  \sum_\alpha \psi_\alpha\otimes\psi_\alpha^\dagger\cdot  B^{(\mu)}}^{-N_-}
\!\!. 
\end{equation} 

We can use again the theorem of section \ref{sec:theorem} obtaining:
\begin{equation}
Z_{pq}\propto \int_{\hat{Gl}(n_b |n_f)} d\mu_H(U)\, \Sdet{U}^{\nu}
e^{-\Str{\frac{N_+}{N} MU}}  
\Sdet{1+\sigma  \frac{1}{N_+} M\cdot U^{-1} +\frac{2}{N} B^{(\mu)} U 
B^{(\mu)} U^{-1}}^{-N_-}\ .
\end{equation} 
This result is again exact for any finite $N$. If we now take the large-$N$
limit while keeping $M$ and $B^{(\mu)}$ fixed\footnote{In the theory with real
chemical potential in the next section this limit is called weak
non-Hermiticity limit 
\cite{Fyodorov:1996sx}. While we inherit
the same scaling here our operators are always Hermitian.} 
we obtain finally:
\begin{equation}
\lim_{N\to\infty}Z_{pq}=
\int_{\hat{Gl}(n_b |n_f)} d\mu_H(U)\, \Sdet{U}^{\nu}
e^{-\sqrt{\frac{\sigma}{2}}\Str{M\cdot U+ M\cdot U^{-1}}-\Str{ B^{(\mu)} U
    B^{(\mu)} U^{-1}}} \nonumber  
\end{equation} 
where our only approximation has been the expansion of the superdeterminant. 

This expression is equivalent to the $\epsilon\chi$PT 
effective partition
\cite{Damgaard0604054}, where in order to match 
we use the following relations \cite{Osborn:2006xf,Akemann:2007rf} 
\begin{equation}
\label{mapping_constants}
\begin{split}
m_i N\sqrt{\frac{\sigma}{2}}=&m_i\frac{\Sigma V}{2} \ ,\\
\mu_i \sqrt{N}=
&\mu_{D_i} F_\pi  \sqrt{V}\ .
\end{split}
\end{equation} 
Here $\mu_D$ is a dimensional constant, instead of the $\mu$ used above that is
dimensionless. The additional parameter $F_\pi$ is the pion decay constant,
the second low energy constant in the leading order chiral Lagrangian.
Thus we have established the $\epsilon\chi$PT-chRMT equivalence for
all $k$-point correlation functions with imaginary chemical potential.

\section{Real chemical potential\label{sec:real_chem}}
Before starting the proof for real chemical potential it's better to spend a 
few words on the definition of theories with bosonic quarks. It may help
understanding both the result and the proof. 

\subsection{Bosonic partition function with $ \mu\neq 0
  $\label{sec:bosonic_z_mu}}
The sign problem makes it impossible to write the inverse
Dirac operator as a convergent bosonic integral: Hermitisation is needed
\cite{Feinberg:1997dk,Janik:1996xm} and a new particle content of the theory
comes out \cite{Toublan:1999hx,Splittorff:2003cu,KimJac06}. 

\subparagraph*{}
Let's consider the simplest example, 
$ n_b=1 $, $ n_f=0 $, 
following \cite{KimJac06}. The
partition function is given by: 
\begin{equation}
\left< \frac{1}{\Det{D+\mu\gamma_0+m}} \right>\ .
\end{equation}
In the microscopic theory (QCD) the inverse power of the determinant should be
drawn from the integration of bosonic quarks. This integration is well
defined only if the integral is convergent, that is,  
if the Hermitian part of the operator is positive (or negative) defined;
this requirement is fulfilled when Dirac operator is given by a non-Hermitian
(kinetic) part plus a mass term (proportional to the identity operator), but
this is not the case for a Dirac operator suffering from the sign problem. 

The way to bypass this problem is to add an additional bosonic conjugate quark
and a fermionic one with mass $ -m^* $ and chemical potential $ -\mu $,
respectively, 
\begin{eqnarray}
\left< \frac{1}{\Det{D+\mu\gamma_0+m}} \right>&=&
\left<
\frac{\Det{D-\mu\gamma_0-m^*}}{\Det{D+\mu\gamma_0+m}\Det{D-\mu\gamma_0-m^*}}
\right>\nonumber\\ 
&\propto &	
\left< \frac{\Det{D-\mu\gamma_0-m^*}}{\Det{
{\bmm 0 & D+\mu\gamma_0+m \\ D-\mu\gamma_0-m^* &0\emm}}} \right>\ .
\end{eqnarray}
The operator in the denominator is now anti-Hermitian and hence has only pure
imaginary eigenvalues. If we consider the matrix 
\begin{equation}
\bpm \varepsilon & D+\mu\gamma_0+m \\ D-\mu\gamma_0-m^* & \varepsilon\epm
\label{anti_herm}
\end{equation}
(sending $ \varepsilon\rightarrow 0 $ at the end) we have an operator
fulfilling the positivity requirements to have a properly defined theory. This
process is usually called Hermitisation \cite{Feinberg:1997dk,Janik:1996xm},
and although it could seem as 
just a mathematical trick this is not the case: the
possibility of writing the inverse determinants as convergent integrals is a
feature of the underlying theory. 
This procedure modifies the form of the mass term in $\epsilon\chi$PT 
\cite{Splittorff:2003cu,KimJac06}, and for the most general result we refer to 
eq. (\ref{echPTmaster}),
where the generalisation to any number $ n_b $ of bosonic uncoupled quarks is
straightforward: $ (N_b|0)\rightarrow (N_b+N_b^*|N_b^*) $. 
Despite this regularisation is a feature of the microscopic underlying
theory, it may happen that we have, in an effective theory, properly 
defined integrals even without it, and that no such Hermitisation is needed. 
This is the case of non-Hermitian Random
Matrix Theory with uncoupled bosons \cite{Akemann:2004zu,AOSV,KimJac06}. 

\subsection{Equivalence for real chemical potential\label{sec:realmu}}
As already hinted in sect. \ref{sec:resolvent} and explained in sect.
\ref{sec:bosonic_z_mu} the resolvent method may require the Hermitisation of
the boson-boson part of Dirac operator \cite{Toublan:1999hx}: 
\begin{eqnarray}
Z_{pq}&=&\left\langle \dfrac{\prod_i^{n_v}\Det{D_i+z_i}}
{\prod_i^{n_v}\Det{D_i+\tilde z_i}} \prod_f^{N_f}\Det{D_f+m_f} \right\rangle \\
&=&\left\langle \dfrac{\prod_i^{n_v}\Det{D_i+z_i}\Det{D_i+\tilde z_i}^*}
{\prod_i^{n_v}\Det{D_i+\tilde z_i}\Det{D_i+\tilde z_i}^*}
\prod_f^{N_f}\Det{D_f+m_f} \right\rangle  \nonumber\\ 
&=&\lim_{\varepsilon\rightarrow 0} \left\langle
\dfrac{\prod_i^{n_v}\Det{D_i+z_i}\Det{D_i+\tilde z_i}^*} 
{\prod_i^{n_v}\Det{(D_i+\tilde z_i)(D_i^\dagger+\tilde z_i^*)+\varepsilon^2}}
\prod_f^{N_f}\Det{D_f+m_f} \right\rangle .  \nonumber 
\end{eqnarray} 

The first line is the proper definition of the resolvent, the second, that is
obtained from the first by a trivial step, is the Hermitised version, the last
line is the regularised integral. Hermitisation is necessary both for the
computation below and for the underlying microscopic theory, but it's not
always necessary for making the computation in RMT
\cite{Akemann:2004zu,KimJac06}.   

In the second line the additional valence quarks are in conjugate pairs $
(n_b+n_b^*|n_v+n_v^*+N_f) $ (the replacement $\tilde z_i\rightarrow z_i$ in the
numerator is irrelevant). It will be sufficient to consider the conjugate
fermionic quarks as independent quarks with mass $ -m^* $ and chemical
potential $ -\mu $; concerning the bosonic conjugate quark, as already seen in
sect. \ref{sec:bosonic_z_mu}, there's a deep reason that forces 
us to treat them simultaneously. 

We mention in passing that a 
purely bosonic theory with bosons in conjugated pairs $(n_b+n_b^*,0)$ occurs
when applying the replica trick or Toda lattice equation
\cite{Stephanov:1996ki,AOSV,Splittorff:2003cu,KimJac06}. 
\medskip 

In the following we will prove the equivalence of the partition functions
of $\epsilon\chi$PT and chRMT at non zero chemical potential with $ n_f $
fermionic quarks with given masses and baryonic potential $ (m_f,\mu_f) $, and
$ n_b $ couples of conjugated bosonic quarks, with parameters $ (m_b,\mu_b) $
and $ (-m_b^*,-\mu_b) $. The partially quenched theory with $ N_f $ physical
quarks, $ n_v $ couples of conjugated fermionic quarks and $ n_v $ couples of
bosonic quarks will result as a special case. The equivalence of the spectra
follows applying the resolvent method in both theories. 

We will again use the the two-matrix model as was introduced by Osborn
\cite{Osborn:2004rf}, but 
the same calculation can be done also for the Stephanov model. 
The equivalent of the QCD 
Dirac operator for a quark with mass $m_f$ and chemical 
potential $\mu_f$ is given by: 
\begin{equation}
{\mathcal D}_f+m_f\equiv \bpm m_f \ID_{N_+} & iA+\mu_f B \\ iA^\dagger +\mu_f
B^\dagger & 
m_f\ID_{N_-}\epm \ .
\end{equation} 
where $A$ and $B$ are complex $N_+\times N_-$ random matrices.

The partition function equivalent to partially quenched QCD is:
\begin{equation}
Z_{pq}=\int dA \ dB\ w(A)\ w(B)\ \frac{\prod_{f}^{n_f}\Det{\bmm m_f \ID_{N_+}
    & iA+\mu_f B \\ iA^\dagger +\mu_f B^\dagger & m_f\ID_{N_-}\emm}} 
{\prod_{b}^{n_b}\Det{\bmm m_b \ID_{N_+} & iA+\mu_b B \\ iA^\dagger +\mu_b
    B^\dagger & m_b\ID_{N_-}\emm}\Det{\bmm -m_b^* \ID_{N_+} & iA-\mu_b B \\
    iA^\dagger -\mu_b B^\dagger & -m_b^*\ID_{N_-}\emm}} \ ,
\end{equation} 
with a Gaussian weight function 
\begin{equation}
w(X)=\Exp{-\sigma N \Tr{X^\dagger X}}\ .
\end{equation} 

In order to write the inverse determinants as bosonic Gaussian integrals we
perform the same anti-Hermitisa\-tion as in eq. (\ref{anti_herm}), obtaining an
anti-Hermitian matrix apart from an $ \varepsilon $ times the identity. The
regularised denominator is: 
\begin{eqnarray}
&&(-)^{N_+-N_-}\Det{\bpm m_b \ID_{N_+} & iA+\mu_b B \\ iA^\dagger 
+\mu_b B^\dagger & 
    m_b\ID_{N_-}\epm \cdot \bpm -m_b^* \ID_{N_+} & iA-\mu_b B \\ iA^\dagger
    -\mu_b B^\dagger & -m_b^*\ID_{N_-}\epm-\varepsilon^2 \ID_{N_++N_-}}=
  \nonumber \\  
&&=\Det{\bpm \varepsilon \ID_{N_+}& 0 &m_b \ID_{N_+} & iA+\mu_b B \\ 0 &
    \varepsilon \ID_{N_-} &iA^\dagger +\mu_b B^\dagger & m_b\ID_{N_-}\\ -m_b^*
    \ID_{N_+} & iA-\mu_b B & \varepsilon\ID_{N_+} & 0 \\ iA^\dagger -\mu_b
    B^\dagger & -m_b^*\ID_{N_-} & 0 & \varepsilon \ID_{N_-}\epm} \\ 
&&=\Det{\bpm
\epsilon +\frac{m-m^*}{2} & i A & \frac{m+m^*}{2} & \mu B  \\
i A^\dagger & \epsilon +\frac{m-m^*}{2} & \mu B^\dagger & \frac{m+m^*}{2} \\
-\frac{m+m^*}{2} & -\mu B & \epsilon  -\frac{m-m^*}{2} & -i A \\
-\mu B^\dagger & -\frac{m+m^*}{2}  & -i A^\dagger & \epsilon -\frac{m-m^*}{2}  
\epm}. \nonumber
\end{eqnarray}

We introduce two sets of $N_+$ and $N_-$ complex supervectors in $(2 n_b|n_f)$
to write the ratio of determinants as a Gaussian integral  
\begin{eqnarray} 
\label{part_anti_herm}
Z_{pq}&=&\int dA\ dB\ d(\psi,\psi^*,\phi,\phi^*) \Exp{-\sigma N
  \left(\Tr{A^\dagger A}+\Tr{B^\dagger B}\right)}  \\ 
&&\times\ \Exp{-\sum_{f}^{n_f}  
\bpm \psi^{*}_{f,\alpha} \\ \phi^{*}_{f,\beta}\epm 
\bpm  m_f\ID_{\alpha,\alpha^\prime} &  \left(i A+\mu_f
  B\right)_{\alpha,\beta^\prime} \\ 
\left(i A^\dagger+ \mu_f B^\dagger\right)_{\beta,\alpha^\prime} &
  m_f\ID_{\beta,\beta^\prime}\epm  
\bpm \psi_{f,\alpha^\prime} \\ \phi_{f,\beta^\prime}\epm}  \nonumber\\ 
&&\times\ \Exp{-\sum_{b}^{n_b}  
\bpm \psi^{*}_{2b-1,\alpha} \\ \phi^{*}_{2b-1,\beta} \\ \psi^{*}_{2b,\alpha}
  \\ \phi^{*}_{2b,\beta}\epm  
\bpm \epsilon +\frac{m_b-m_b^*}{2} & i A_{\alpha\beta^\prime} &
  \frac{m_b+m_b^*}{2} & \mu_b B_{\alpha\beta^\prime} \\ 
i A_{\beta\alpha^\prime}^\dagger & \epsilon +\frac{m_b-m_b^*}{2} & \mu_b
  B_{\beta\alpha^\prime}^\dagger & \frac{m_b+m_b^*}{2} \\ 
-\frac{m_b+m_b^*}{2} & -\mu_b B_{\alpha\beta^\prime} & \epsilon
  -\frac{m_b-m_b^*}{2} & -i A_{\alpha\beta^\prime} \\ 
-\mu_b^* B_{\beta\alpha^\prime}^\dagger & -\frac{m_b+m_b^*}{2}  & -i
  A_{\beta\alpha^\prime}^\dagger & \epsilon -\frac{m_b-m_b^*}{2}  \epm 
\bpm \psi_{2b-1,\alpha^\prime} \\ \phi_{2b-1,\beta^\prime} \\
  \psi_{2b,\alpha^\prime} \\ \phi_{2b,\beta^\prime}\epm}. \nonumber 
\end{eqnarray} 

As before we have reduced all the dependence on random matrices in terms of
Gaussian functions: 
\begin{eqnarray}
&&\Exp{-\sigma N \ A_{\alpha,\beta}A_{\alpha,\beta}^* -i A_{\alpha,\beta} 
\left( \sum_f \psi_{f,\alpha}^*\phi_{f,\beta}+\sum_b\left (
\psi_{2b-1,\alpha}^*\phi_{2b-1,\beta}  -\psi_{2b,\alpha}^*\phi_{2b,\beta}
\right ) \right) \right. \nonumber \\ 
&&\left. -iA_{\alpha,\beta}^*  \left( \sum_f
\phi_{f,\beta}^*\psi_{f,\alpha}+\sum_b\left (
\phi_{2b-1,\beta}^*\psi_{2b-1,\alpha}  -\phi_{2b,\beta}^*\psi_{2b,\alpha}
\right)\right) } \\ 
&&\times\ \Exp{-\sigma N \ B_{\alpha,\beta}B_{\alpha,\beta}^* -
    B_{\alpha,\beta}  
\left( \sum_f \mu_f \psi_{f,\alpha}^*\phi_{f,\beta}+\sum_b \mu_b \left (
\psi_{2b-1,\alpha}^*\phi_{2b,\beta}  -\psi_{2b,\alpha}^*\phi_{2b-1,\beta}
\right ) \right) \right. \nonumber \\ 
&&\left. -B_{\alpha,\beta}^*  \left( \sum_f \mu_f
\phi_{f,\beta}^*\psi_{f,\alpha}+\sum_b \mu_b \left (
-\phi_{2b,\beta}^*\psi_{2b-1,\alpha}  +\phi_{2b-1,\beta}^*\psi_{2b,\alpha}
\right )\right) }\ , \nonumber 
\end{eqnarray} 
and we can perform the Gaussian integration completing the squares:
\begin{eqnarray}
&&\Exp{-\frac{1}{\sigma N} \sum_{\alpha,\beta}{\psi^*_{g,\alpha}\Gamma^A_{gl}
      \phi_{l,\beta} \phi^*_{m,\beta}
      \Gamma^A_{m,n}\psi_{n,\alpha}}+\frac{2}{\sigma N^2}
    \sum_{\alpha,\beta}{\psi^*_{g,\alpha}\Gamma^B_{gl} \phi_{l,\beta}
      \phi^*_{m,\beta} \Gamma^B_{m,n}\psi_{n,\alpha}}}= \\ \nonumber 
&=&\Exp{-\frac{1}{\sigma N} \Str{\Gamma^A \cdot \sum_\alpha \left( \phi_\alpha
      \otimes \phi_\alpha^\dagger \right) \cdot \Gamma^A \cdot \sum_\beta \left(
      \psi_\beta \otimes \psi_\beta^\dagger \right) - \frac{2}{N} \Gamma^B \cdot 
\sum_\alpha  \left( \phi_\alpha \otimes \phi_\alpha^\dagger \right) \cdot \Gamma^B
       \cdot \sum_\beta \left( \psi_\beta \otimes \psi_\beta^\dagger \right)}}\ . 
\end{eqnarray}
Here we have introduced the $(2n_b|n_f)\times(2n_b|n_f)$ supermatrices 
\begin{equation}
\Gamma^A=\bpm \left. \bmm 1 & 0 \\ 0 & -1\emm\right\}\times n_b & 0 \\ 0 &
      {\bf 1}_{n_f} \epm \ ,\ \  
\Gamma^B=\sqrt{\frac{N}{2}}
\bpm \left. \bmm 0& \mu_b \\ -\mu_b &0 \emm \right\} \times n_b & 0
      \\ 0 & \left. \mu_f \right\}\times n_f\epm \ ,  
\end{equation}
and the mass matrix:
\begin{equation}
M=N \bpm \left.\bmm \varepsilon+\frac{m_b-m_b^*}{2} & \frac{m_b+m_b^*}{2} \\
-\frac{m_b+m_b^*}{2} & \varepsilon-\frac{m_b-m_b^*}{2}\emm \right\}\times n_b
&  0 \\ 0 & \left. m_f\right\}\times n_f\epm  \ ,
\end{equation}
anticipating their $N$-dependence below. 
We can rewrite the partition function (\ref{part_anti_herm}):
\begin{eqnarray}
Z_{pq}&=& \int d\left ( \psi,\psi^*\phi,\phi^*\right )  
\Exp{-\Str{\frac{1}{N}M\cdot
    \sum_\beta \psi_\beta \otimes \psi_\beta^\dagger }} \\ 
\nonumber
&&\times\ \Exp{-\Str{\sum_\alpha \phi_\alpha \otimes \phi_\alpha^\dagger \cdot
    \left( \frac{1}{N}M 
+\frac{1}{\sigma N} \Gamma^A \cdot \sum_\beta \psi_\beta \otimes
    \psi_\beta^\dagger  \cdot \Gamma^A  -\frac{2}{\sigma N^2} \Gamma^B \cdot \sum_\beta
    \psi_\beta \otimes \psi_\beta^\dagger \cdot \Gamma^B \right)}} .
\end{eqnarray} 
From this point on the procedure is the same as before: we 
integrate explicitly
the sets of supervectors $ \phi,\phi^* $ and 
use the superbosonisation theorem
\begin{eqnarray}
Z_{pq}&=& \int d\left ( \psi,\psi^*\right )  \Exp{-\Str{\frac{1}{N}
    M\cdot \sum_\beta \psi_\beta \otimes \psi_\beta^\dagger }} \\ 
\nonumber
&&\times\ \Sdet{ \frac{1}{N} M +\frac{1}{\sigma N} \Gamma^A \sum_\beta
  \psi_\beta \otimes \psi_\beta^\dagger  \Gamma^A  -\frac{2}{\sigma N^2}
  \Gamma^B \sum_\beta  \psi_\beta \otimes \psi_\beta^\dagger
  \Gamma^B }^{-N_-} \\ 
&\propto&\int_{\hat{Gl}(2n_b |n_f)}d\mu_H(U) \Sdet{U}^{N_+} \, e^{-\Str{
    \frac{N_+}{N} M\cdot U}} 
\Sdet{ \frac{1}{ N} M +\frac{N_+}{\sigma N}\, \Gamma^A \cdot U\cdot  \Gamma^A
    -\frac{2N_+}{\sigma N^2} \Gamma^B \cdot U \cdot \Gamma^B }^{-N_-}
\!\!\!\! .
    \nonumber 
\end{eqnarray} 

This result is valid for finite $N$. 
Once performing the $ N\rightarrow \infty $ weak non-Hermiticity limit
\cite{Fyodorov:1996sx}, keeping $M$ fixed as well as 
$\Gamma^B$, the following result is obtained
\begin{equation}
\lim_{N\to\infty}
Z_{pq}=\int_{\hat{Gl}(2n_b |n_f)}\!d\mu_H(U) \Sdet{U}^\nu
\Exp{\Str{-\sqrt{\frac{\sigma}{2}} 
M\left( U+\Gamma^A\cdot U^{-1} \cdot  \Gamma^A\right)+ 
\Gamma^A \Gamma^B U \Gamma^B \Gamma^A U^{-1}}}.
\end{equation}

Rotating all the matrices under the 
superunitary
transformation $ X\rightarrow T^\dagger\cdot X
\cdot T $ with 
\begin{equation}
T=\bpm \left. \bmm i/\sqrt{2} & 1/\sqrt{2} \\ -i/\sqrt{2} &
1/\sqrt{2}\emm\right\}\times n_b & 0 \\ 0 &  {\bf 1}_{n_f} \epm ,
\end{equation}
we obtain the new result for Hermitised $\epsilon\chi$PT with real $\mu\neq0$,
generalising previous results in the literature
\cite{Splittorff:2003cu,KimJac06}:
\begin{equation}
\lim_{N\to\infty}
Z_{pq}=\int_{\hat{Gl}(2n_b |n_f)}d\mu_H(U) \Sdet{U}^\nu
\Exp{\Str{-\sqrt{\frac{\sigma}{2}} 
 \hat{M}\left( U+ I \cdot U^{-1} \cdot I\right)+
 B_+^{(\mu)} U B_-^{(\mu)} U^{-1}}},
\label{echPTmaster}
\end{equation}
where
\begin{eqnarray}
&&\hat{M}\equiv T^\dagger\cdot M\cdot T=
\bpm \left. \bmm \varepsilon N & -i m_b N \\ -i m_b^* N & \varepsilon N
\emm\right\}\times n_b & 0 \\ 0 & \left. m_f N \right\}\times n_f \epm
\nonumber \\ 
&& B_+^{(\mu)}\equiv T^\dagger \cdot \Gamma^A \cdot \Gamma^B\cdot T= 
\sqrt{\frac{N}{2}}\bpm \left. \bmm -\mu_b & 0 \\ 0 & \mu_b \emm\right\}\times n_b & 0 \\ 0 & \left. \mu_f  \right\}\times n_f\epm  
\nonumber \\
&& B_-^{(\mu)}\equiv T^\dagger \cdot \Gamma^B \cdot \Gamma^A\cdot T=
\sqrt{\frac{N}{2}}\bpm \left. \bmm \mu_b & 0 \\ 0 & -\mu_b \emm\right\}\times n_b & 0 \\ 0 & \left. \mu_f  \right\}\times n_f\epm  
\nonumber \\
&& I\equiv T^\dagger \cdot \Gamma^A \cdot T=
\bpm \left. \bmm 0 & -i \\ i & 0\emm\right\}\times n_b & 0 \\ 0 & {\bf
  1}_{n_f}\epm .
\end{eqnarray}
It is expressed in term of $\Sigma  $,
$ F_\pi $ and chemical potential using eq. (\ref{mapping_constants}).
For details on the Haar measure $d\mu_H(U)$ we refer to the next section.

The expression above is a generalisation 
of $\epsilon \chi $PT
with one pair of bosons \cite{Splittorff:2003cu,KimJac06}, and of 
$\epsilon\chi $PT with $N_f$
fermions
$ \epsilon\chi $PT \cite{Verbaarschot:2005rj}. 
The signature of the boson-boson block in the metric $I$ differs from 
\cite{Splittorff:2003cu} by a factor of $i$ because we have used 
an anti-Hermitisation here, see subsection \ref{sec:bosonic_z_mu}, 
instead of a Hermitisation there. 
The main point is that the two mass terms have different signs if 
considering bosonic or fermionic quarks, see eqs. (102) and (131) of [24],
respectively.

The particular cases of eq. (\ref{echPTmaster}) already known in 
literature \cite{Splittorff:2003cu,KimJac06,Verbaarschot:2005rj} 
have been derived starting from
the symmetries of the microscopic theory under vector and axial 
transformations too, 
and we suppose that the same arguments
can be applied to the general case.
The existence of two different
matrices $ B^{(\mu)}_\pm $  
is due to the fact that the covariant derivative has a different
behaviour on  bosonic and fermionic quarks in 
$\epsilon \chi $PT as explained in \cite{KimJac06}. 
\\ 

\section{Superbosonisation theorem\label{sec:theorem}}
The aim of this section is to find a way to express integrals of functions of
external products of supervectors in terms of integrals over a smaller
space. In formulas we are going to study 
\begin{equation}
\int d(\psi,\psi^*) \ f\left ( \sum_k \psi_k\otimes \psi_k^\dagger\right ),
\end{equation}
where $ \psi_k,\psi^\dagger_k $ are complex vectors of $ n_b $ bosons and $
n_f $ fermions, $ d(\psi,\psi^*) $  is a shorthand notation for $ \prod_k^N
d\psi_k \ d\psi_k^*$ and $f$ is a function defined over $(n_b|n_f)$
supermatrices. 

The main idea is to embed $ \sum_k \psi_k\otimes \psi_k^\dagger $ in some
manifold where we can define a $\delta$-function and where computations are
feasible. At a merely symbolic level we have: 
\begin{equation}
\label{rho_idea}
\int d(\psi,\psi^*) \ f\left ( \sum_k \psi_k\otimes \psi_k^\dagger\right )= 
\int_{\cal M} da\, f(a) \int d(\psi,\psi^*) \ 
\delta_{\cal M}\left (a- \sum_k \psi_k\otimes
\psi_k^\dagger\right ) \ .
\end{equation}
In order to give a meaning to the equation above we have to specify which is
the manifold $ {\cal M} $ 
we use and which is the measure we use on it (the $ \delta$-function 
depends on it). The use of a $\delta$-function requires $f$ to be
continuous with respect to the metric chosen. 

A good choice is to take as $ {\cal M}$ in eq. (\ref{rho_idea}) 
the manifold of super-Hermitian matrices.  
One could ask whether or not a $\delta$-function may be used dealing with
matrices like external product of supervectors \cite{BEKYZ}; in appendix
\ref{sec:app_delta} we show that such a $\delta$ may be properly defined.  
The $\delta$-function may be written using an extension of the usual
Fourier-transform representation of the $\delta$-function on Hermitian
matrices (see appendix \ref{sec:app_delta}). 

A crucial ingredient in this proof will be the possibility to flip the order
of integration. Changing the order between commuting and anti-commuting
variables is not a problem, but problems can arise when considering the case of
two commuting variables. In order to see where these problems come
from, and how to avoid them, let's consider a simple example, the one of a
single commuting number vector of length $1$: 
\begin{eqnarray}
\int_{\mathbb C}d^2z\, f(z\cdot z^*)&=&\int_{\mathbb C}d^2z \, \int_{\mathbb
  R}dx\, f(x) \, \delta(x-zz^*) \nonumber \\ 
&=&\int_{\mathbb R}dx\, f(x) \int_{\mathbb C}d^2z \, \delta(x-zz^*) \\
&=&\int_{\mathbb R}dx\, f(x) \int_{\mathbb C}d^2z \,\int_{\mathbb R}dy
  \, e^{iy(x-zz^*)} \ .\nonumber 
\end{eqnarray}

The first change of variable is always allowed when considering converging
integrals. At this point we need to flip the order of integration of $y$ and
$z,z^*$, and this is an illicit step since the integrals are not
converging. This problem may be avoided considering a real quantity $ \eta>0 $ 
\begin{eqnarray}
\int_{\mathbb C}d^2z\, f(z z^*)&=&\int_{\mathbb C}d^2z\, f(z z^*)
e^{(\eta-\eta)zz^*} \nonumber \\ 
&=&\int_{\mathbb R}dx\, f(x) e^{\eta x}\int_{\mathbb C}d^2z \,\int_{\mathbb
  R}\frac{dy}{2\pi}\, e^{iy(x-zz^*)} e^{-\eta zz^*} \nonumber \\ 
&=&\int_{\mathbb R}dx\, f(x) e^{\eta x} \,\int_{\mathbb R}\frac{dy
  }{2\pi}\, e^{iyx} \int_{\mathbb C}d^2z e^{-(\eta +iy)zz^*}
\nonumber \\ 
&=&\int_{\mathbb R}dx\, f(x) e^{\eta x} \,\int_{\mathbb R}\frac{dy
  }{2\pi}\, e^{iyx} \frac{(-i)\pi}{y-i\eta} \nonumber \\ 
&=&\int_{\mathbb R}dx\, f(x) e^{\eta x} \, e^{-\eta x} \pi \, \theta(x)=
\pi\int_0^\infty dx\,  f(x) \ .
\end{eqnarray}
and all the steps are mathematically rigorous any time we can apply a 
$\delta$-distribution 
as a functional to the function $ f(x) e^{\eta x} $. The symbol
$\theta(x)$ indicates the step function.  

This easy example is conceptually not too different from the proof of the
following theorem.

\begin{theorem}\emph{(Superbosonisation)} \label{theo_equivlance}
Let $f$ be a 
function defined on the  $(n_b|n_f)$ supermatrices, then the following
identity holds 
\begin{equation}
\int \prod_k^N d\psi_k d\psi_k^*\, f\left(\sum_k \psi_k \otimes \psi_k^\dagger
\right)  \propto \ \int_{\hat{Gl}(n_b|n_f)} d\mu_H(U)\ \Sdet{U}^{N} \ f(U)
\label{theorem_eq} 
\end{equation} 
whenever the integral on the l.h.s. is well defined. The first integration is
performed over $N$ complex supervectors in $(n_b|n_f)$ and $d\mu_H(U)$ denotes
the Haar measure over $\hat{Gl}(n_b|n_f)$. 
\end{theorem}

\paragraph{}
The idea of expressing integrals of a function of an external product $\sum_k
\psi_k \otimes \psi_k^\dagger$ in terms of an integral of the same function
over a simpler space has been widely used in physics:
it was derived for the first time in \cite{Kawamoto:1981hw} for external 
products of Grassmannian vectors appearing in exponential function
(the term ``bosonisation'' comes from this kind of application), subsequently
in \cite{Fyodorov:2002ub} for commuting number vectors and in 
\cite{Hackenbroich:1995,LSSS,Berruto:2001ty,Efetov:1999,Guhr:2006ee}
for supervectors.
This (super)bosonisation may be seen as application of the Riesz
Representation Theorem\footnote{This theorem ensures that a bounded linear
  functional of functions over a locally compact Hausdorff space may be
  computed like an integration of that function over that space using a proper
  measure \cite{RudinBook}.}. However, up to our knowledge, a graded version
of the last theorem is lacking. 

By coincidence the superbosonisation theorem has been independently developed
at the very same time\footnote{This work was already presented by one of the 
authors in ``QCD in extreme conditions'', Frascati 6-8 August 2007.} 
of an analogous theorem on superbosonisation by
Littelmann, Sommers and Zirnbauer \cite{LSZ,BEKYZ}. Anyway the proof they give
is different from ours, theirs is an algebraic proof, ours uses
analysis instruments. The main difference is that we base our proof on the
existence of a $\delta$-function 
fulfilling eq. (\ref{rho_idea}), and that collaboration 
develops a powerful apparatus in order not to use such an equation
\cite{LSZ}. For this reason we have chosen to show in appendix
\ref{sec:app_delta} all the details concerning the mathematical rigorousness
of our definition.

\begin{proof}

As already hinted above we write the l.h.s. of eq. (\ref{theorem_eq})
introducing an additional integration over super-Hermitian matrices: 

\begin{eqnarray}
\int d(\psi,\psi^*)\, f\left ( \sum_k\psi_k\otimes\psi_k^\dagger \right )
&=&
\int_{H=H^\dagger} dH\, f(H) \, e^{\eta\Str{H}} \int d(\psi,\psi^*)\,
\delta\left ( \sum_k\psi_k\otimes\psi_k^\dagger -H\right ) \, e^{-\eta\sum_k
  \psi^\dagger_k\cdot \psi_k} \nonumber \\ 
&\propto& \label{proof_th_ft}
\int_{H=H^\dagger} dH\, f(H) \, e^{\eta\Str{H}} \int d(\psi,\psi^*)\,
e^{-\eta\sum_k \psi_k^\dagger\cdot \psi_k} \int_{F=F^\dagger} dF\, e^{iF\left
  ( H- \sum_k\psi_k\otimes\psi_k^\dagger \right )}  
\nonumber\\
&=& \int_{H=H^\dagger} dH\, f(H) \, e^{\eta\Str{H}} \int_{F=F^\dagger} dF\,
e^{iFH}\, \Sdet{\eta+iF}^{-N} 
\end{eqnarray}
where $d(\psi,\psi^*)\equiv\prod_k^N d\psi_k d\psi_k^*$ and $d\psi_k
d\psi_k^*=\prod_b^{n_b} d^2_{\mathbb C}\psi_{k,b}\, \prod_f^{n_f}d\psi_{k,f}
d\psi_{k,f}^*$. We denote by $\psi^\dagger=\psi^{*\, T}$, and 
$\psi\otimes\psi^\dagger$ is the
external product and $\psi^\dagger\cdot\psi$ the scalar product in
$(n_b|n_f)$. The measure $dH$ is the flat measure in the independent entries
of the super-Hermitian matrix. 

From now on let's focus on the second integral. This quantity reminds of a
quantity already computed in \cite{Guhr:2006ee}, unfortunately we cannot use
this result because it does not take care about the boundary terms arising
from the diagonalisation of super-Hermitian matrices (Efetov-Wegner terms). In
the following we will never change the $Z$-gradings of the integration
manifold, and hence no boundary terms will arise \cite{Rothstein}. 

Writing $ F $ and $ H $ in terms of blocks, $dF=dF_1\, dF_2\, d\Phi\,
d\Phi^\dagger$ 
\begin{equation}
F=\bpm F_1 & \Phi^\dagger \\ \Phi & F_2 \epm \ ,\ \ \ \ 
H=\bpm H_1 & \Theta^\dagger \\ \Theta & H_2 \epm\ ,
\end{equation}
we obtain:
\begin{eqnarray}
{\mathcal I}&\equiv&\int_{F=F^\dagger} dF\, e^{iFH}\, \Sdet{F-i\eta}^{-N} \\
&=&
\int dF_1\, dF_2\, d\Phi\,d\Phi^\dagger\ \left(\frac{ \Det{F_2-i\eta - \Phi
    (F_1-i\eta)^{-1} \Phi^\dagger}}{\Det{F_1-i\eta}}\right)^N\!  
\Exp{i\Tr{ F_1 H_1 + \Phi^\dagger \Theta}-i\Tr{ H_2 F_2 +\Phi \Theta^\dagger
}}. \nonumber 
\end{eqnarray} 

As a first step we note that if we diagonalise the matrix $F_1$ and make
an analytic continuation in the eigenvalues we see that the only poles in
the expression above are the ones where at least one eigenvalue of $F_1$ is
$i\eta$. Considering the integration on the eigenvalues as a complex contour
integral and closing the integral in the upper semicircle (or lower one,
depending on the signature of the matrix $H_{1}$ \footnote{The matrices $H_1$
  and $H_2$ are complex number Hermitian matrices.}) we have non vanishing
contributions
only if every path of the eigenvalues winds the pole in $i\eta$. We can
consider this integral as a contour integral around $ i\eta $ whenever
$\theta(H_{1})>0$, otherwise it's zero.  
Since there are no poles apart from the ones in $i\eta$ we can fix the contour
integral as we prefer. We choose the modulus of the eigenvalues as equal to
$1$, hence we have that the matrix $F_1-i\eta$ may be analytically continued
obtaining a unitary matrix.

\begin{eqnarray}
{\mathcal I}&=&\theta(H_{1}) e^{-\eta \Tr{H_1}} \oint_{U(n_b)} dF_1 \int
dF_2\, d\Phi\,d\Phi^\dagger\ \left(\frac{ \Det{F_2-i\eta - \Phi F_1^{-1}
    \Phi^\dagger}}{\Det{F_1}}\right)^N \nonumber\\ 
&&\times\ \Exp{i\left[ F_1 H_1 + \Phi^\dagger \Theta \right]-i\left[ H_2 F_2
    +\Phi \Theta^\dagger \right]} \ . 
\end{eqnarray} 

The notation $ \oint_{U(n_b)}dF_1$ stands for an integral over the manifold
$U(n_b)$ considering as a measure the analytic continuation of the flat
measure of Hermitian matrices. It is related to the Haar measure $d\mu_U$
integration by: 
\begin{equation}
 \oint_{U(n_b)}dF_1=\int_{U(n_b)} d\mu_U(F_1) \, \Det{F_1}^{n_b}\ .
 \label{analytic_cont_haar} 
\end{equation}

The relation between these two integrals is the same as between the two
integrals below describing a circuitation around zero, with $z=e^{i\theta}$: 
\begin{equation}
\int_{0}^{2\pi}d\theta=\oint_{|z|=1} dz\,\frac{1}{z} \neq \oint_{|z|=1}dz.
\end{equation}

Eq. (\ref{analytic_cont_haar}) can be derived diagonalising the matrix $ F_1=U
\cdot \f\cdot U^\dagger $ where $\f_j=e^{i\theta_j}$: 
\begin{eqnarray}
 \oint_{U(n_b)}dF_1 &\equiv& \prod_i \oint d\f_i \Delta^2(\{\f\}) \int dU 
\nn\\
&=&
\prod_i \oint \frac{d\f_i}{\f_i} \Det{\f}\ \Delta(\{\f\}) \Delta\left
(\{\f^{-1}\}\right ) \Det{\f}^{n_b-1} \int dU  \nonumber\\ 
&=&
\prod_i \int_0^{2\pi} d\theta_i \left | \Delta(\{e^{i\theta}\}) \right |^2
\Det{\f}^{n_b} \int dU 
\nn\\
&\equiv&
\int_{U(n_b)} d\mu_U(F_1) \, \Det{F_1}^{n_b} \ .
\end{eqnarray}

We consider now that
\begin{eqnarray}
\int_{A=A^\dagger}dA\ \Det{A}^N e^{-i\Tr{A H_2}} 
&=&\int_{A=A^\dagger}dA\ \Det{A-i\eta}^N e^{-i\Tr{(A-i\eta) H_2}} \nn\\
&=&\int_{A=A^\dagger}dA\ \Det{A-i\eta- \Phi^\dagger F_1^{-1} \Phi}^N
e^{-i\Tr{(A-i\eta- \Phi F_1^{-1} \Phi^\dagger) H_2}} \ .
\label{proof_th_invar} 
\end{eqnarray} 

The first equality comes from the analyticity of the integrand in the diagonal
entries of the matrix
and the second comes from the analogue of the contour
invariance in superanalysis \cite{DeWitt1984}, applied to the real and
imaginary parts of the Hermitian matrix entries. Applying this equivalence we
obtain 
\begin{eqnarray}
{\mathcal I}&=&\theta(H_{1}) e^{\eta\Tr{H_2}-\eta\Tr{H_1}} \oint_{U(n_b)} dF_1
\int dF_2\, d\Phi\,d\Phi^\dagger\ \left(\frac{ \Det{F_2}}{\Det{F_1}}\right)^N
\nonumber\\ 
&&\times\ \Exp{i\Tr{ F_1 H_1 +\Phi^\dagger \Theta}-i\Tr{ H_2 F_2 + \Phi
    \Theta^\dagger }-i\Tr{ H_2 \Phi F_1^{-1} \Phi^\dagger }}\ .
\label{proof_th_1} 
\end{eqnarray} 

The subsequent step is to perform the $\Phi,\Phi^\dagger$ integration, using
\begin{eqnarray}
\nonumber
\Tr{ \Phi^\dagger \Theta}-\Tr{\Phi \Theta^\dagger + H_2 \Phi F_1^{-1}
  \Phi^\dagger} &=& 
-\Tr{ \Theta \Phi^\dagger + \Phi \Theta^\dagger + H_2 \Phi F_1^{-1}
  \Phi^\dagger}
\nn\\
&=&-\Tr{ \Theta F_1 F_1^{-1}\Phi^\dagger + H_2 \Phi \Theta^\dagger H_2^{-1}+
  H_2  
\Phi F_1^{-1} \Phi^\dagger}\nn\\
&=&-\Tr{ \left(\Theta F_1 +H_2 \Phi \right) \cdot 
\left( F_1^{-1}\Phi^\dagger +\Theta^\dagger H_2^{-1} \right)-\Theta F_1
\Theta^\dagger H_2^{-1}}. 
\end{eqnarray} 

We can transform the fermionic variables of integration
$\Phi\rightarrow\tilde\Phi \equiv  H_2 \Phi+\Theta F_1 $ and
$\Phi^\dagger\rightarrow \tilde \Phi^\dagger \equiv F_1^{-1}
\Phi^\dagger+\Theta^\dagger H_2^{-1}$ 
and thus perform the Gaussian integration,
\begin{eqnarray}
\int d\Phi\,d\Phi^\dagger \Exp{-i\Tr{ H_2 \Phi  F_1^{-1}\Phi^\dagger
    }+i\Tr{\Theta F_1 \Theta^\dagger H_2^{-1}}} 
= \frac{\Det{H_2}^{n_b}}{\Det{F_1}^{n_f}} e^{i\Tr{\Theta F_1 \Theta^\dagger
    H_2^{-1}}}\int d\tilde\Phi\,d\tilde\Phi^\dagger 
e^{-i\Tr{ \tilde\Phi
    \tilde\Phi^\dagger }} ,
\end{eqnarray}
where the last integration is just a constant. Getting back to
eq. (\ref{proof_th_1}) we have: 
\begin{eqnarray}
{\mathcal I}&\propto&\theta(H_{1}) \Det{H_2}^{n_b}\,
e^{-\eta\Str{H}}\oint_{U(n_b)} 
dF_1 \int 
dF_2\,  \frac{ \Det{F_2}^{N}}{\Det{F_1}^{N+n_f}} 
e^{i\Tr{ F_1 H_1 }-i\Tr{ H_2 F_2 }+i\Tr{ \Theta F_1 \Theta^\dagger
    H_2^{-1} }}  
\label{proof_th_2} \\
&=&\theta(H_{1}) \Det{H_2}^{n_b}\, e^{-\eta\Str{H}} \int_{F_2=F_2^\dagger}
dF_2\,   \Det{F_2}^{N} e^{-i\Tr {H_2 F_2}}
\oint_{U(n_b)} \! dF_1 \frac{ 1}{\Det{F_1}^{N+n_f}} e^{i\left[ F_1
    \left (H_1-\Theta^\dagger H_2^{-1} \Theta \right ) \right]}.
\nonumber
\end{eqnarray} 

Let us consider again the whole eq. (\ref{proof_th_ft}). We can now perform
the integration in $ F_1 $ and $ F_2 $, using the integrals in appendix
\ref{sec:app_proof_int}
\begin{eqnarray} 
\int d(\psi,\psi^*)\, f\!\left ( \sum_k\psi_k\otimes\psi_k^\dagger \right )
&\propto&
\int_{H_1=H_1^\dagger}dH_1\, \theta(H_{1}) \int_{H_2=H_2^\dagger}dH_2\,
  \Det{H_2}^{n_b}\int  d\Theta\, d\Theta^\dagger\ f(H)\nonumber\\ 
&&\times\ 
\int_{F_2=F_2^\dagger} \!dF_2 \Det{F_2}^{N} e^{-i\Tr {H_2 F_2}}
\oint_{U(n_b)} \!\!\!dF_1 \frac{ 1}{\Det{F_1}^{N+n_f}}
e^{i\left[ F_1
  \left (H_1-\Theta^\dagger H_2^{-1} \Theta\right ) \right]} \nonumber\\ 
&\propto&\int_{H_1=H_1^\dagger}dH_1\, \theta(H_{1})
  \int_{H_2=H_2^\dagger}dH_2\, \Det{H_2}^{n_b}\, \int 
d\Theta\, d\Theta^\dagger\ f(H)
  \nonumber\\ 
&&\times\int_{F_2=F_2^\dagger} dF_2\,   \Det{F_2}^{N} e^{-i\Tr
  {H_2 F_2}} 
 \Det{H_1-\Theta^\dagger H_2^{-1} \Theta}^{N+n_f-n_b}  \nonumber\\
&\propto&\int_{H_1=H_1^\dagger} dH_1\, \theta(H_{1}) \oint_{U(n_f)} dH_2 
\int d\Theta\, d\Theta^\dagger\,  \frac{\Det{H_1-\Theta^\dagger H_2^{-1}
  \Theta}^{N+n_f-n_b}}{\Det{H_2}^{N+n_f-n_b}}f(H)   \nonumber \\ 
&=& \int_{H_1=H_1^\dagger} dH_1\, \theta(H_{1}) \oint_{U(n_f)} dH_2 
\int d\Theta\, d\Theta^\dagger\,   \Sdet{\bmm H_1 & \Theta^\dagger \\ \Theta &
  H_2 \emm}^{N+n_f-n_b}f(H)\ . 
\label{proof_th_3}
\end{eqnarray} 

The last step missing is to find a relation between the measure used in the
equation above and the Haar measure. This relation is: 
\be
\label{claim_equivalence}
\int_{H_1=H_1^\dagger} dH_1\, \theta(H_1) \oint_{U(n_f)} dH_2 
\int d\Theta\, d\Theta^\dagger\,   \Sdet{\bmm H_1 & \Theta^\dagger \\ \Theta &
  H_2 \emm}^{n_f-n_b} 
= \int_{\hat{Gl}(n_b |n_f)} d\mu_H\left( \bmm H_1 & \Theta^\dagger \\ \Theta &
H_2 \emm\right). 
\ee
It is proved in the following.
The manifolds we are integrating are the same, in fact the
$H_{1}=H_{1}^\dagger>0$ is equivalent to the boson-boson \cite{Dalmazi:2000bs}
base manifold of $\hat{Gl}(n_b |n_f)$ that is $Gl(n_b)/U(n_b)$
\cite{Zirnbauer:1998,Damgaard:1998xy}. The analytic continuation of the
eigenvalues of $H_2=H_2^\dagger$ making a circuitation around zero is
equivalent to the fermion-fermion part $U(n_f)$ \cite{Zirnbauer:1998}. 

We have now to consider the integration measures in
eq. (\ref{claim_equivalence}). 
The point is that the measure on the l.h.s. is  the Haar measure of
super-Hermitian matrices (a group closed under addition) computed on the
analytically continued manifold. It is induced by the flat metric: 
\begin{equation}
\Str{dU\cdot dU}\ ,
\end{equation} 
but the Haar measure on the r.h.s. is the analytical continuation of the Haar
measure on super-Unitary matrices
\cite{Zirnbauer:1998,Dalmazi:2000bs,Damgaard:1998xy} (a group closed under
multiplication), induced by the metric \cite{Damgaard:1998xy}: 
\begin{equation}
\Str{dU\cdot dU^{-1}}=\Str{U^{-1} dU\cdot U^{-1} dU}.
\end{equation} 

In order to find a relation between these two matrices one has to compute the
Berezinean of the transformation \cite{DeWitt1984,Zirnbauer:1998}: 
\begin{equation}
{\mathfrak B}=\SDet{\frac{(U^{-1}\, dU)_{ij}}{dU_{mn}}}{ij,mn}.
\end{equation} 

Since this is not a change of variables (the differentials $dU$ are the same)
and hence no Z-gradings are changed, no boundary (Efetov-Wegner) 
terms arise in the
superintegration. 
We write this Berezinean matrix in the block structure (Latin indices stay for
bosons and Greek for fermions): 
\begin{eqnarray}
\frac{U^{-1}\, dU}{dU}&=&
\bpm bb/bb & bb/ff & bb/fb & bb/bf \\ 
ff/bb & ff/ff & ff/fb & ff/bf \\ 
fb/bb & fb/ff & fb/fb & fb/bf \\ 
bf/bb & bf/ff & bf/fb & bf/bf \epm 
\nonumber\\ 
&=& \bpm
U_{ab}^{-1} \otimes {\bf 1}_{n_b} & 0 & U_{a\beta}^{-1} \otimes {\bf 1}_{n_b}
& 0\\ 
0 & U_{\alpha\beta}^{-1} \otimes {\bf 1}_{n_f} & 0 & U_{\alpha b}^{-1} \otimes
{\bf 1}_{n_f}  \\ 
U_{\alpha b}^{-1} \otimes {\bf 1}_{n_b} & 0 & U_{\alpha\beta}^{-1} \otimes
{\bf 1}_{n_b} & 0 \\ 
0 & U_{a\beta}^{-1} \otimes {\bf 1}_{n_f} & 0 & U_{a b}^{-1} \otimes {\bf
  1}_{n_f}  
\epm\ .
\end{eqnarray} 
Computing the superdeterminant of the matrix above we obtain
\begin{equation}
{\mathfrak B}=\SDet{\frac{(U^{-1}\,
    dU)_{ij}}{dU_{mn}}}{ij,mn}=\Sdet{U}^{n_f-n_b}. 
\end{equation} 

As a consequence we have:
\begin{equation}
d\mu_H\left( \bmm H_1 & \Theta^\dagger \\ \Theta & H_2 \emm\right) = dH_1\,
dH_2\, d\Theta\, d\Theta^\dagger\,   \Sdet{\bmm H_1 & \Theta^\dagger \\ \Theta
  & H_2 \emm}^{n_f-n_b}  \ ,
\end{equation} 
that is eq. (\ref{claim_equivalence}). Together with eq. (\ref{proof_th_3}) we
obtain the proof of our theorem eq. (\ref{theorem_eq}). \qed 
\end{proof}

\section{Conclusions}
In the present paper we prove that $\chi$PT in the $\varepsilon$-regime 
and chRMT have the very same spectral properties in the
microscopic (weak non-Hermiticity) limit. The equivalence holds for
zero, imaginary and real chemical potentials. 
This result is achieved by proving the equivalence (up to an irrelevant overall
constant) of the partition function of the partially quenched chRMT, with
$n_b$ bosonic and $n_f$ fermionic quarks, to an integral defined over the
maximal Riemannian subspace of the linear group of $(n_b|n_f)$ supermatrices
for any finite $N$. Performing the $N\rightarrow\infty$ limit of this integral
one obtains the 
partially quenched $\epsilon\chi$PT.
The one obtained for real chemical potential was derived here 
for the first time and agrees with the previously
known purely fermionic and purely bosonic cases. 
An essential ingredient of
this computation is a superbosonisation theorem connecting integrals of
external products of supervectors to integrals over $\hat{Gl}(n_b|n_f)$. An
explicit parametrisation of integrals over $\hat{Gl}(n_b|n_f)$ is provided
as well.

The equivalence we have established by matching the generating functionals 
of density correlations holds also in cases where the densities are not yet
known explicitly. Only for at most 
two different chemical potentials all densities
have been derived so far. 
Our results also shed some light on universality in chRMT with chemical
potential by matching the Gaussian one- and two-matrix model of Stephanov and
Osborn, respectively. This gives hope to establish a further reaching
universality beyond Gaussian RMT.

\indent

\noindent
\underline{Acknowledgements}:

We would like to thank 
P.~H.~Damgaard, A.~Patella, K.~Splittorff, B.~Svetitsky and
J.~J.~Verbaarschot for useful discussions,
M. Zirnbauer for communications and sharing an unpublished paper, and the 
CPT Luminy for hospitality, where part of this work was being done.
This work was supported by 
EU network ENRAGE MRTN-CT-2004-005616 (G.A.),
EPSRC grant EP/D031613/1 and 
Brunel University BRIEF award no. 707.

\appendix
\section{Defining 
$ \delta $-functions on super-Hermitian matrices\label{sec:app_delta}}
The question is whether or not one can define a $\delta$-distribution such that
\begin{equation}
f\left(\sum_k \psi_k\otimes\psi_k^\dagger \right)=\int dM \, f(M)
\delta\left(M-\sum_k \psi_k\otimes\psi_k^\dagger\right). 
\label{app:delta_eq_1}
\end{equation} 

In order to clarify the idea we take the same simple but nontrivial
example as in \cite{BEKYZ}, that is $k=1$, $(n_b|n_f)=(0|2)$. The external
product is a $2\times 2$ matrix, it's terms are only nilpotent commuting
numbers $\psi_i$ and $\psi_j^*$ for $i,j=1,2$. We can consider this matrix as 
belonging to a superanalytic continuation of a manifold of real dimension 4
(at least). We could take as a such manifold $U(2)$, $Gl(2)/U(2)$, Hermitian
matrices or real matrices. Despite in the rest of the work we have used 
Hermitian matrices, for this simple example we will use the one with the
simplest notation, real matrices: 
\begin{equation}
a\equiv \bpm a_{11} & a_{12} \\ a_{21} & a_{22} \epm \equiv
\psi\otimes\psi^\dagger = 
\bpm \psi_1\psi_1^* & \psi_1\psi_2^* \\ \psi_2\psi_1^* & \psi_2\psi_2^* \epm\ .
\end{equation} 

Once considering the external product of vectors like an element of a commuting
number manifold, the well definiteness of the $\delta$-distribution is
inherited from the one of the base manifold by superanalyticity in the real
parameters of the manifold. In formulas: 
\begin{equation}
f(a)\equiv f(a_{11},a_{12},a_{21},a_{22})=\int dm\, \delta(m-a) \,
f(m_{11},m_{12},m_{21},m_{22}) \ ,
\end{equation} 
where the integration manifold is the base manifold (real matrices) $\int dm
\equiv \int_{-\infty}^{\infty}\prod_{i,j=1}^{2}dm_{ij}$, and the $\delta$
distribution is the one of the base manifold too, $\delta(m-a)\equiv
\prod_{i,j=1}^{2}\delta(m_{ij}-a_{ij})$.  

The last thing to be verified is that the superanalytic continuation of the
$\delta$-distribution is properly defined; this was already shown in
\cite{DeWitt1984}. Given a quantity $x=x_B+x_S$, where the first part 
is the body of the number, and the second is the nilpotent part. Given 
a function with a
sufficient number of derivatives in $x_B$ such that
$f(x_B+x_S)=\sum_{n=0}^\infty f^{(n)}(x_B) \frac{x_S^n}{n!}$ is properly
defined, we have:  
\begin{eqnarray}
\int_{-\infty}^{+\infty}dy \, \delta(y-x_B-x_S)\, f(y)&\equiv& 
\int_{-\infty}^{+\infty}dy \, \sum_{n=0}^{\infty}
\delta^{(n)}(y-x_B)\frac{(-x_S)^n}{n!} \, f(y) \nonumber \\ 
&=&\sum_{n=0}^{\infty} \frac{(x_S)^n}{n!} \int_{-\infty}^{+\infty}dy \,
\delta(y-x_B) \, f^{(n)}(x_B) = f(x_B+x_S)  \ .
\end{eqnarray} 

Completing the discussion on the example above we can write explicitly
$\delta(m-a)$: 
\begin{equation}
\delta(m-a)=\delta(m)-\sum_{i,j=1}^2 \psi_i\psi_j^* \partial_{m_{ij}}
\delta(m) + \psi_1\psi_1^*\psi_2\psi_2^*
(\partial_{m_{11}}\partial_{m_{22}}+\partial_{m_{12}}\partial_{m_{21}})
\delta(m) 
\end{equation} 
where $m$ is a real number $2\times 2$ matrix.

\subparagraph{}
The generalisation for what is done above to eq. (\ref{app:delta_eq_1}) is
straightforward: one considers the external product as an element belonging to
the superanalytic continuation of a manifold. The manifold we choose is the
one of super-Hermitian matrices: the boson-boson and the fermion-fermion block
are Hermitian matrices, and the $\delta$ is defined as the superanalytic
continuation of the product of $\delta$'s over the real and imaginary parts of
the independent elements of the matrices; boson-fermion and fermion-boson
blocks are made of independent Grassmann integration variables and $\delta$
may be represented as: 
\begin{equation}
\delta(\theta-\tilde \theta)\propto (\theta-\tilde \theta) \propto 
\int d\xi \, e^{i \xi (\theta-\tilde \theta)}.
\end{equation} 

As an exponential representation of $\delta$ is allowed both for commuting and
anticommuting variables, such a representation of $\delta$-function on
super-Hermitian matrices is allowed too: 
\begin{eqnarray}
f(a)&=&\int_{M=M^\dagger}dM\, f(M)\, \delta(M-a) \\ \nonumber
&\propto& \int_{M=M^\dagger}dM\, f(M)\, \int_{F=F^\dagger}dF\, e^{i\Str{
    F(M-a)}} \, 
\end{eqnarray} 
where $dM$ and $dF$ stand for the flat measure over the independent entries of
super-Hermitian matrices: 
\begin{equation}
dM\equiv  \, \prod_{i=1}^{n_b}\prod_{\alpha=1}^{n_f}\, dM_{i,\alpha}\,
dM_{\alpha,i}\, \prod_{i=1}^{n_b} dM_{i,i}\,  \prod_{i>j} d{\mathbb R}e\,
M_{i,j}\, d{\mathbb I}m\, M_{i,j}  
\prod_{\alpha=1}^{n_f} dM_{\alpha,\alpha}\,  \prod_{\alpha,\beta} d{\mathbb
  R}e\, M_{\alpha,\beta}\, d{\mathbb I}m\, M_{\alpha,\beta} \ .
\end{equation}

\section{The integrals in eq. (\ref{proof_th_2})\label{sec:app_proof_int}}

\subsection{Boson-boson block} 
Let's start from 
\begin{eqnarray}
&&
\oint_{U(n_b)} dF_1\, 
\frac{ 1}{\Det{F_1}^{N+n_f}}\Exp{i\Tr{ F_1 \tilde H }} \ =
\nn\\
&&=
\int_{U(n_b)} d\mu_U (F_1)\, \Det{F_1}^{-N-n_f+n_b} \Exp{i\Tr{ F_1 \tilde H }}
\nonumber \\ 
&&=\Det{\tilde H}^{N+n_f-n_b} \int_{U(n_b)} d\mu_U (F_1)\, \Det{F_1 \tilde
  H}^{-N-n_f+n_b} \Exp{i\Tr{ F_1 \tilde H }}\ , 
\end{eqnarray} 
where for simplicity of notation we use $\tilde H=H_1-\Theta H_2^{-1}
\Theta^\dagger$. The integral in the last line may be performed using the
character expansion. This is a particular case of a more general integral
computed in \cite{Schlittgen:2002tj}; the result is: 
\begin{equation}
\sum_r \frac{\alpha_r^{(0)}}{d_r} \alpha_r^{(N+n_f-n_b)} \chi_r(0)
\end{equation} 
where the sum is over the irreducible representations of $Gl(n_b)$. The
quantity $\chi_r(0)$ is zero for all the representation apart
from the trivial one. The result is just a constant\footnote{In order to
  compute the exact value one has just to substitute the relation
$\alpha_r^{(\nu)}=\det_{ij}\left[ \frac{1}{n_j-\nu+i-j} \right]$,
where $r=(n_1,\dots,n_b)$ are the labels of the representation.}.
\begin{equation}
\oint_{U(n_b)} dF_1 \frac{ 1}{\Det{F_1}^{N+n_f}} e^{i\Tr{ F_1 \tilde H
    }} \propto  
\Det{\tilde H}^{N+n_f-n_b}  \label{app_proof_int_1}.
\end{equation} 

This result, together with the argument of analytic continuation of Hermitian
matrices, gives an alternative way for computing the Ingham-Siegel integral as
was done in \cite{Fyodorov:2002ub}.

\subsection{Fermion-fermion block}
The fermion-fermion block integral has already been computed by Guhr
\cite{Guhr:2006ee}. The result is: 
\begin{equation}
\int_{F_2=F_2^\dagger} dF_2\, \Det{F_2}^{N} \Exp{-i\Tr{ F_2 H_2 }} \propto
\prod_j\delta^{(N+n_f-1)}(h_j) \ .
\end{equation} 
Here we don't have any problem with possible Efetov-Wegner terms arising from
these diagonalisation: both $F_2$ and $H_2$ are complex Hermitian matrices and
no Z-grading will be changed during diagonalisation \cite{Rothstein}. 

If we integrate this functional with a given function analytic in the matrix
entries, using the Cauchy integral formula we obtain: 
\begin{eqnarray}
\int d\mu_U(U)\, \prod_j\int dh_j\, \Delta(\{h\})^2\,  g(U h U^\dagger) \
\prod_j\delta^{(N+n_f-1)}(h_j)  
&\propto&\int d\mu_U(U) \oint \prod_j dh_j\frac{1}{h_j^{N+n_f}}   
\, \Delta(\{h\})^2\,  g(U h U^\dagger) 
\nn\\ 
&=&\oint_{U(N_f)}dH_2\ \frac{1}{\Det{H_2}^{N+n_f}} \ g(H_2)\ .
\end{eqnarray}

\end{document}